\documentclass[
    ,final            
    ,numberedheadings 
  ]
  {aipproc}

\layoutstyle{6x9}

\input epsf
\let\sec=\section

\def\japitem#1{\smallskip\noindent\rlap{#1}\hglue1.8em\hangindent1.8em}
\def\enditem{\smallskip\noindent}

\def\japref{\parskip=0pt\par\noindent\hangindent\parindent
    \parskip =2ex plus .5ex minus .1ex}

\def\gs{\mathrel{\lower0.6ex\hbox{$\buildrel {\textstyle >}
 \over {\scriptstyle \sim}$}}}
\def\ls{\mathrel{\lower0.6ex\hbox{$\buildrel {\textstyle <}
 \over {\scriptstyle \sim}$}}}
\newcount\japequationnum
\global\japequationnum=0
\def\bookdisp#1$${\leftline{\hfill{$\displaystyle#1$}
    \global\advance\japequationnum by 1
    \hfill (\the\japequationnum )}$$}
\everydisplay{\bookdisp}
\def\japsub{\rm\scriptscriptstyle}

\def\kms{{\,\rm km\,s^{-1}}}

\def\hompc{{\,h\,\rm Mpc^{-1}}}
\def\mpcoh{{\,h^{-1}\,\rm Mpc}}

\def\japsub{\rm\scriptscriptstyle}

\newcommand{\plotter}[2]{\centering \leavevmode \epsfxsize=#2\textwidth \epsfbox{#1}\medskip}
\newcommand{\japplottwo}[2]{\centering \leavevmode 
\epsfxsize=0.43\textwidth \epsfbox{#1}
\hglue 1em
\epsfxsize=0.43\textwidth \epsfbox{#2}
\medskip}
\newcommand{\japplottwob}[2]{\centering \leavevmode 
\epsfxsize=0.49\textwidth \epsfbox{#1}
\hglue 1em
\epsfxsize=0.49\textwidth \epsfbox{#2}
\medskip}

\def\m@th{\mathsurround=0pt }
\def\eqalign#1{\null\,\vcenter{\openup1\jot \m@th
 \ialign{\strut\hfil$\displaystyle{##}$&$\displaystyle{{}##}$\hfil
 \crcr#1\crcr}}\,}

\def\topinsert{\begin{figure}[ht]}

\begin{document}

\title{Implications of 2dFGRS results \\ on cosmic structure}

\author{J.A. Peacock$^*$}{
  address={Institute for Astronomy, University of Edinburgh,\\
Royal Observatory, Edinburgh EH9 3HJ, UK}
}

\begin{abstract}
The 2dF Galaxy Redshift Survey is the first to observe more
than 100,000 redshifts, making possible precise measurements of
many aspects of galaxy clustering.
The spatial distribution of galaxies can be studied as a function
of galaxy spectral type, and also of broad-band colour.
Redshift-space distortions are detected with a high degree of
significance, confirming the detailed Kaiser distortion from
large-scale infall velocities, and measuring the distortion parameter
$\beta \equiv \Omega_m^{0.6}/b = 0.49 \pm 0.09$.
The power spectrum is measured to $\ls 10\%$ accuracy for
$k>0.02 \hompc$, and is well fitted by a CDM model with
$\Omega_m h =0.18 \pm 0.02$ and a baryon fraction of $0.17\pm 0.06$.
A joint analysis with CMB data requires $\Omega_m =0.31 \pm 0.05$ and 
$h=0.67\pm0.04$, assuming scalar fluctuations.
The fluctuation amplitude from the CMB is $\sigma_8=0.76 \pm0.04$,
assuming reionization at $z\ls 10$, so that the general level
of galaxy clustering is approximately unbiased, in agreement
with an internal bispectrum analysis.
Luminosity dependence of clustering is however detected at high significance, 
and is well described by a relative bias of $b/b^* = 0.85 + 0.15(L/L^*)$.
This is consistent with the observation that $L^*$ in rich clusters is
brighter than the global value by $0.28 \pm 0.08$ mag.
\end{abstract}

\renewcommand{\thefootnote}{\fnsymbol{footnote}}
\footnotetext[1]{{\sl On behalf of the 2dF Galaxy Redshift Survey team:} Matthew Colless (ANU), 
Ivan Baldry (JHU), Carlton  Baugh (Durham), 
Joss Bland-Hawthorn (AAO), Terry Bridges (AAO),
Russell Cannon (AAO), Shaun Cole (Durham), 
Chris Collins (LJMU), Warrick Couch (UNSW), 
Gavin Dalton (Oxford), Roberto De Propris (UNSW), 
Simon Driver (St Andrews), George Efstathiou (IoA), 
Richard  Ellis (Caltech), Carlos Frenk (Durham), 
Karl Glazebrook (JHU), Carole Jackson (ANU), Ofer Lahav (IoA), Ian Lewis (AAO), 
Stuart Lumsden (Leeds), Steve Maddox (Nottingham), Darren Madgwick (IoA), 
Peder Norberg (Durham), Will Percival (ROE), 
Bruce Peterson (ANU), Will Sutherland (ROE), Keith Taylor (Caltech).}
\renewcommand{\thefootnote}{\arabic{footnote}}

\maketitle

\sec{Aims and design of the 2dFGRS}

The 2dF Galaxy Redshift Survey (2dFGRS) was designed to study the following key
aspects of the large-scale structure in the galaxy distribution:

\japitem{(1)}To measure the galaxy power spectrum $P(k)$ on scales up to a few
hundred Mpc, bridging the gap between the scales of nonlinear
structure and measurements from the the cosmic microwave background (CMB).

\japitem{(2)}To measure the redshift-space distortion of the large-scale clustering
that results from the peculiar velocity field produced by the mass
distribution. 

\japitem{(3)}To measure higher-order clustering statistics in order to
understand biased galaxy formation, and to test
whether the galaxy distribution on large scales
is a Gaussian random field.

\enditem
The survey is designed around the 2dF multi-fibre spectrograph on the
Anglo-Australian Telescope, which is capable of observing up to 400
objects simultaneously over a 2~degree diameter field of view. 
For details of
the instrument and its performance 
see {\tt http://www.aao.gov.au/2df/}, and also
Lewis et~al.\ (2002).

The source catalogue for the survey is a revised and extended version of
the APM galaxy catalogue (Maddox et~al.\ 1990a,b,c); this
includes over 5~million galaxies down to $b_{\japsub J}=20.5$ in both
north and south Galactic hemispheres over a region of almost
$10^4\, {\rm deg}^2$ (bounded approximately by declination $\delta \leq+3^\circ$
and Galactic latitude $b\gs 20^\circ$). 
This catalogue is
based on Automated Plate Measuring machine (APM) scans of 390 plates
from the UK Schmidt Telescope (UKST) Southern Sky Survey. The $b_{\japsub J}$
magnitude system for the Southern Sky Survey is defined by the response
of Kodak IIIaJ emulsion in combination with a GG395 filter,
and is related to the Johnson--Cousins system by $b_{\japsub J} = B -0.304(B-V)$,
where the colour term is estimated from comparison with the SDSS Early
Data Release (Stoughton et al. 2002)
The photometry of the catalogue is calibrated with numerous
CCD sequences and has a precision of approximately 0.15~mag for galaxies
with $b_{\japsub J}=17$--19.5. The star-galaxy separation is as described in
Maddox et~al.\ (1990b), supplemented by visual validation of each galaxy
image.

\begin{figure}[ht]
\plotter{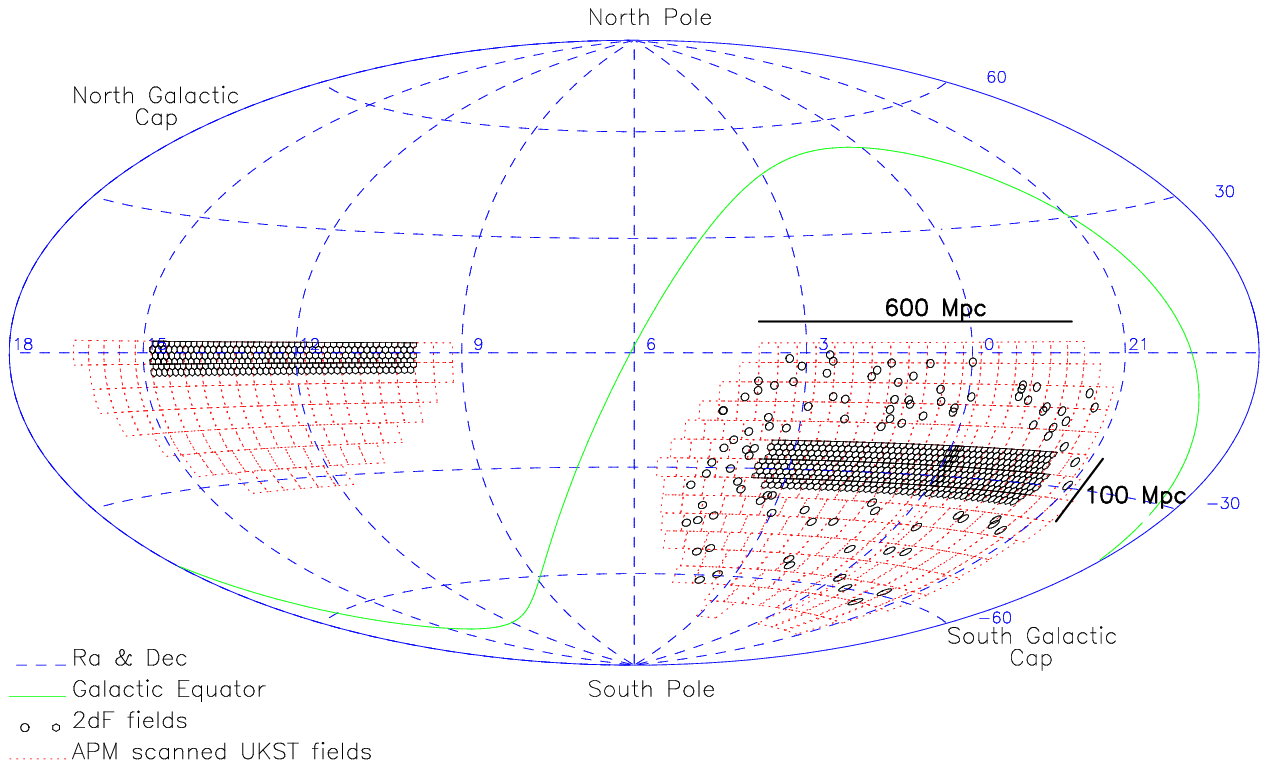}{0.6}
\caption{The 2dFGRS fields (small circles) superimposed on the APM
catalogue area (dotted outlines of Sky Survey plates). There are
approximately 140,000 galaxies in the $75^\circ\times15^\circ$
southern strip centred on the SGP, 70,000 galaxies in the
$75^\circ\times7.5^\circ$ equatorial strip, and 40,000 galaxies in
the 100 randomly-distributed 2dF fields covering the whole area of the
APM catalogue in the south.}
\end{figure}

The survey geometry is shown in Figure~1, and consists of two contiguous
declination strips, plus 100 random 2-degree fields. One strip is in the
southern Galactic hemisphere and covers approximately
75$^\circ$$\times$15$^\circ$ centred close to the SGP at
($\alpha, \delta$)=($01^h$,$-30^\circ$); the other strip is in the northern
Galactic hemisphere and covers $75^\circ \times 7.5^\circ$ centred at
($\alpha, \delta$)=($12.5^h$,$+0^\circ$). The 100 random fields are spread
uniformly over the 7000~deg$^2$ region of the APM catalogue in the
southern Galactic hemisphere. At the median redshift of the survey
($\bar{z}=0.11$), $100\mpcoh$ subtends about 20~degrees, so the two strips
are $375\mpcoh$ long and have widths of $75\mpcoh$ (south) and $37.5\mpcoh$
(north). 

The sample is limited to be brighter than an extinction-corrected
magnitude of $b_{\japsub J}=19.45$ (using the extinction maps of Schlegel et~al.\
1998). This limit gives a good match between the density on the sky of
galaxies and 2dF fibres. Due to clustering, however, the number in a
given field varies considerably. To make efficient use of 2dF, we employ
an adaptive tiling algorithm to cover the survey area with the minimum
number of 2dF fields. With this algorithm we are able to achieve a 93\%
sampling rate with on average fewer than 5\% wasted fibres per field.
Over the whole area of the survey there are in excess of 250,000
galaxies.

\sec{Survey Status}

After an extensive period of commissioning of the 2dF instrument,
2dFGRS observing began in earnest in May 1997, and terminated
in April 2002. 
In total, observations were made of 899 fields,
yielding redshifts and identifications for 232,529 galaxies, 13976 stars
and 172 QSOs, at an overall completeness of 93\%. 
The galaxy redshifts are assigned a quality flag from 1 to 5,
where the probability of error is highest at low $Q$. Most analyses
are restricted to $Q\ge 3$ galaxies, of which there are currently
221,496.
An interim data release took place in July 2001,
consisting of approximately 100,000 galaxies (see Colless et al. 2001
for details). A public release of the full photometric and spectroscopic
database is scheduled for July 2003.

The Colless et al. (2001) paper details the practical steps that
are necessary in order to work with a survey of this sort.
The 2dFGRS does not consist of a simple region sampled with 100\%
efficiency, and it is therefore necessary to use a number of 
masks in order to interpret the data. Two of these concern the
input catalogue: the boundaries of this catalogue, including
`drilled' regions around bright stars where galaxies could not
be detected; also, revisions to the photometric calibration
mean that in practice the survey depth varies slightly with position on the sky.
A futher mask arises from the way in which
the sky is tessellated into 2dF tiles:
near the survey edges and near internal holes,
a lack of overlaps mean that the sampling fraction falls to about 50\%.
Finally, the spectroscopic success rate of each spectroscopic observation
fluctuated according to the observing conditions.
The median redshift yield was approximately 95\%, but with a tail
towards poorer data. The terminal stages of 2dFGRS observing were in
fact devoted to re-observing these fields of low completeness;
nevertheless, approximately 10\% of fields have completeness
lower than 80\%.
This variable sampling makes quantification of the large scale
structure more difficult, particularly for any analysis requiring relatively
uniform contiguous areas. However, the effective survey `mask' can
be measured precisely enough that it can be allowed for in
analyses of the galaxy distribution.

\sec{Galaxy spectra and colours}

Beyond the basic data of positions, magnitudes and redshifts, it is important
on physical grounds to be able to divide the 2dFGRS database into different
categories of galaxies. This has been done in two different ways.
Spectral classification of 2dFGRS galaxies was performed by Folkes et al. (1999)
and Madgwick et al. (2002).
Principal component analysis was used to split galaxies into a superposition
of a small number of templates. Not all of these are robust, owing to
uncalibrated spectral distortions in the 2dF instrument, but it
was possible to derive a robust classification
parameter (termed $\eta$) from the templates, which effectively measures
the emission-line strength (closely related to the star-formation rate).
Galaxies were divided into four spectral classes; their mean spectra
and separate luminosity functions are shown in Figure~2.

\begin{figure}[ht]
\japplottwob{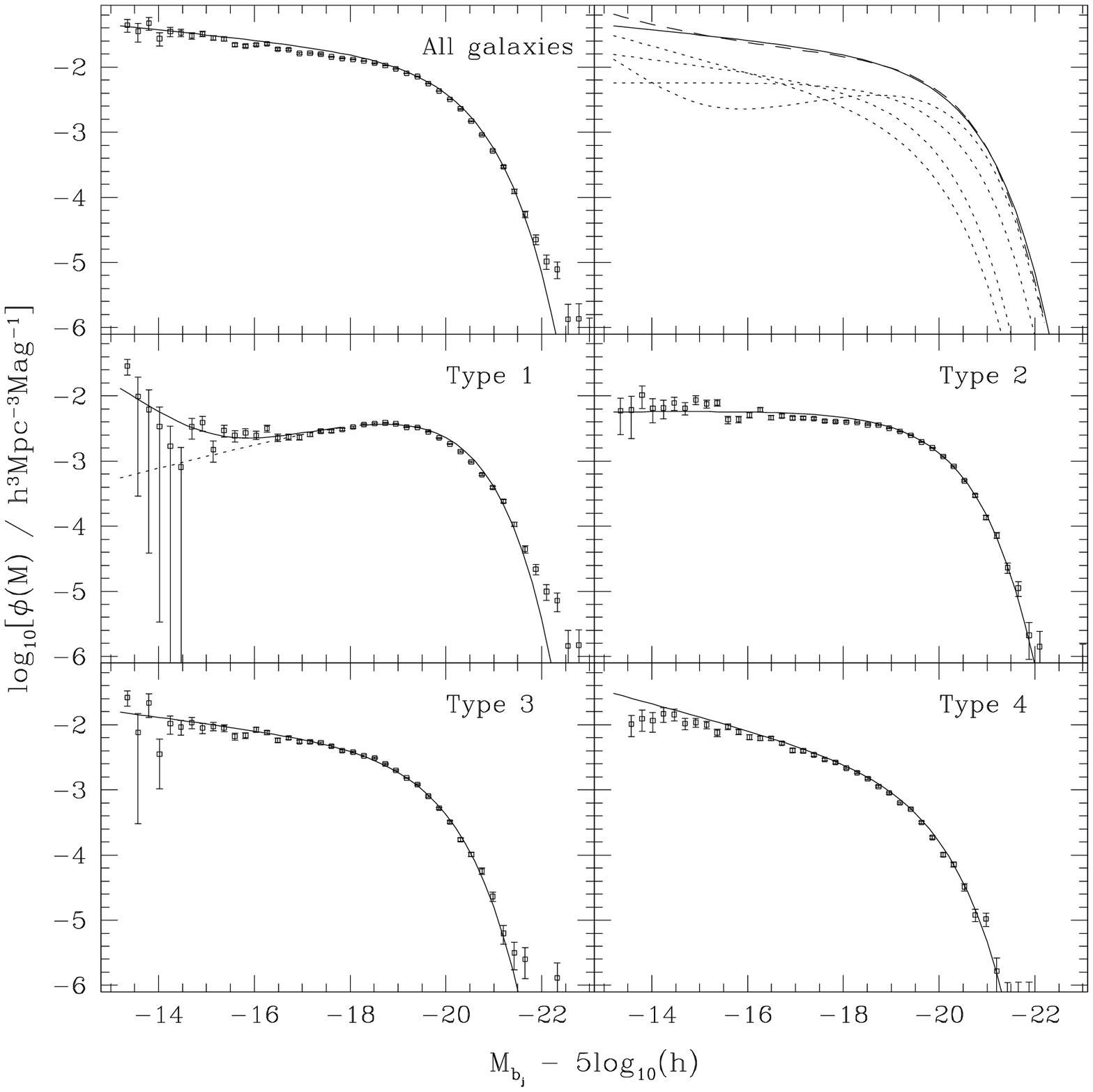}{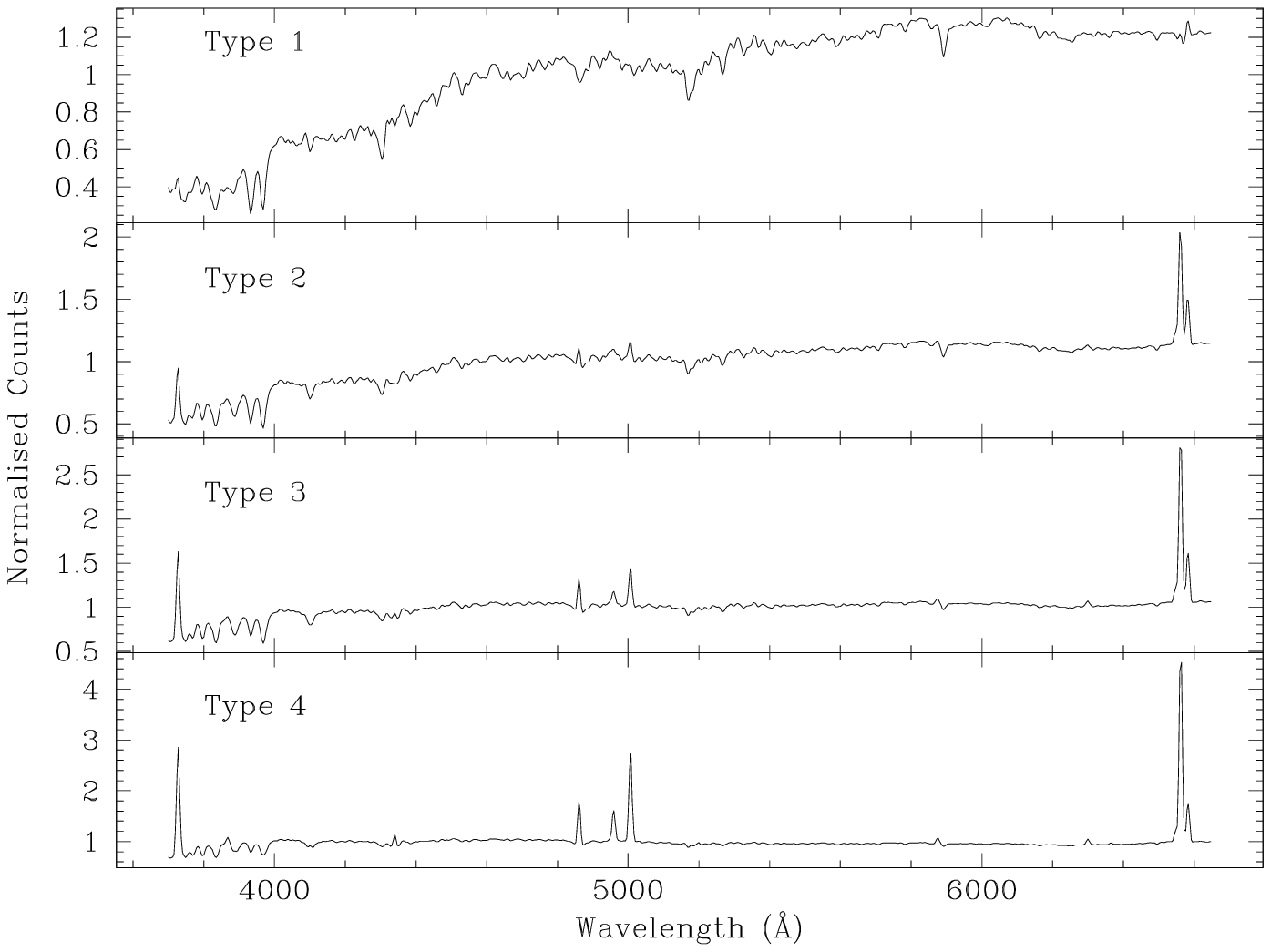}
\caption{
The type-dependent galaxy luminosity function according to Madgwick et al. (2002).
Principal component analysis was used to split galaxies into a superposition
of a small number of templates, and a categorization made based on the
decomposition. Type 1 galaxies are generally E/S0, while later types range
from Sa to Irr.
}
\end{figure}

This classification method has the drawback that it cannot be used beyond
$z=0.15$, where H$\alpha$ is lost from the spectra. Also, the fibres
do not cover the whole galaxy (although Madgwick et al. 2002 show that
aperture corrections are not large in practice).
More recently, we have been able to obtain total broad-band colours
for the 2dFGRS galaxies, using the data from the SuperCOSMOS sky 
surveys (Hambly et al. 2001). These yield $B_{\japsub J}$ from the
same UK Schmidt Plates as used in the original APM survey, but
with improved linearity and smaller random errors (0.09 mag 
rms relative to the SDSS EDR data). The $R_{\japsub F}$
plates are of similar quality, so that we are able to divide
galaxies by colour, with an rms in photographic $B-R$ of about
0.13 mag. The systematic calibration uncertainties are
negligible by comparison, and are at the level of 0.04 mag.
rms in each band. Figure~3 shows that the colour information
divides `passive' galaxies with little active star formation
cleanly from the remainder, uniformly over the whole redshift
range of the 2dFGRS.

\begin{figure}[ht]
\japplottwo{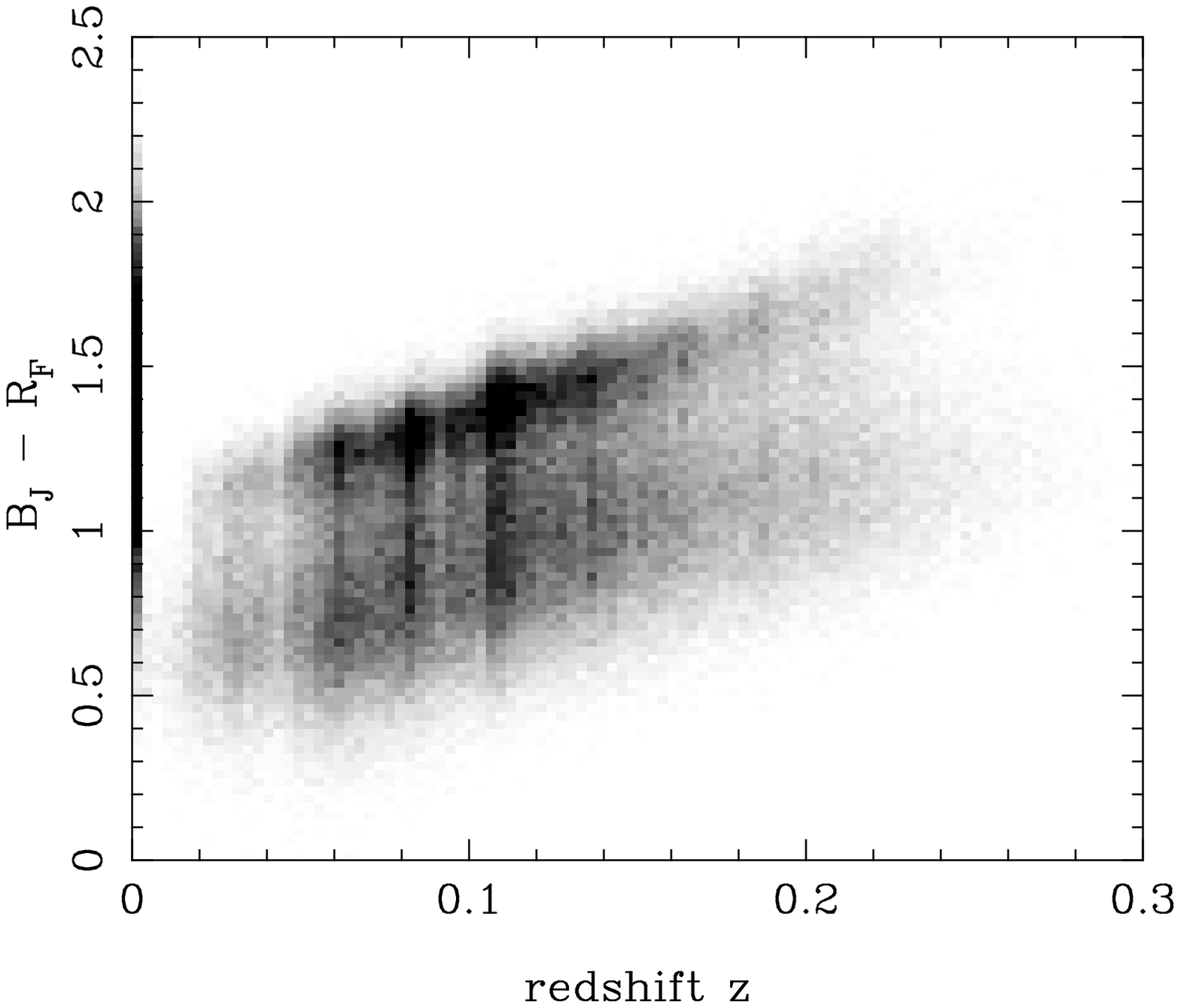}{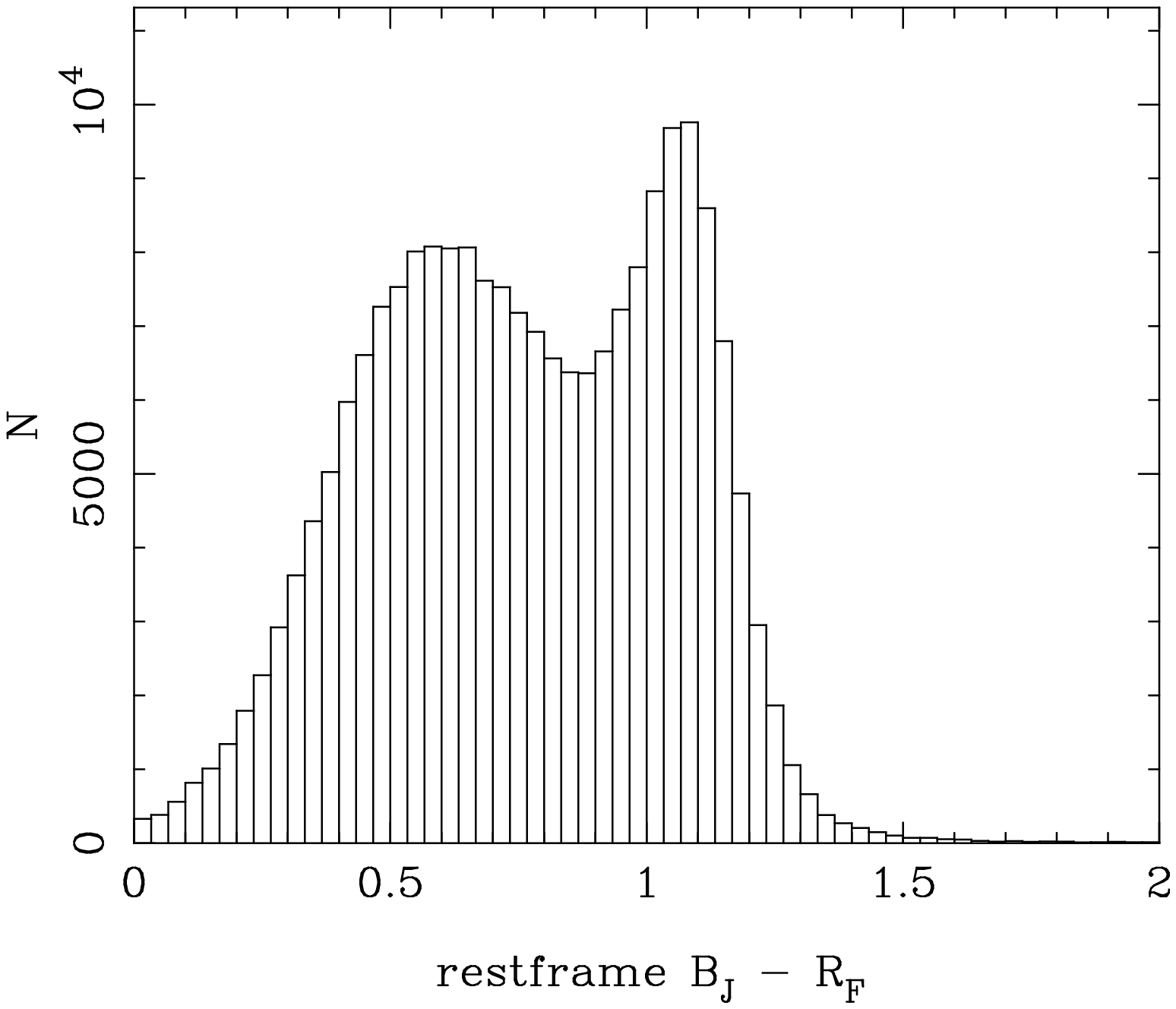}
\caption{
Photographic $B-R$ colour versus redshift for the 2dFGRS. The separation
between `passive' (red) and `active' (blue) galaxies is very clear.
Empirically, $B-R-2.8z$ defines a `restframe' colour whose distribution
is independent of redshift, and very clearly bimodal. This is strongly
reminiscent of the distribution of the spectral type, $\eta$, and we
assume that a division at $(B-R)_0=0.85$ achieves a separation of
`class 1' galaxies from classes 2--4, as was done using spectra by Madgwick et al. (2002).
}
\end{figure}

As an immediate application, we can display the spatial distribution of
2dFGRS galaxies divided according to colour (Figure~4).
The most striking aspect of this image is how closely both sets
of galaxies follow the same structure. The passive subset display
a more skeletal appearance, as expected owing to morphological
segregation of ellipticals. A red-selected survey such as SDSS will
appear more similar to the passive subset of the 2dFGRS, with relatively
low sampling of the more active spectral type 2--4.

\begin{figure}[ht]
\vbox{
\hbox{
\epsfxsize=0.9\textwidth
\epsfbox{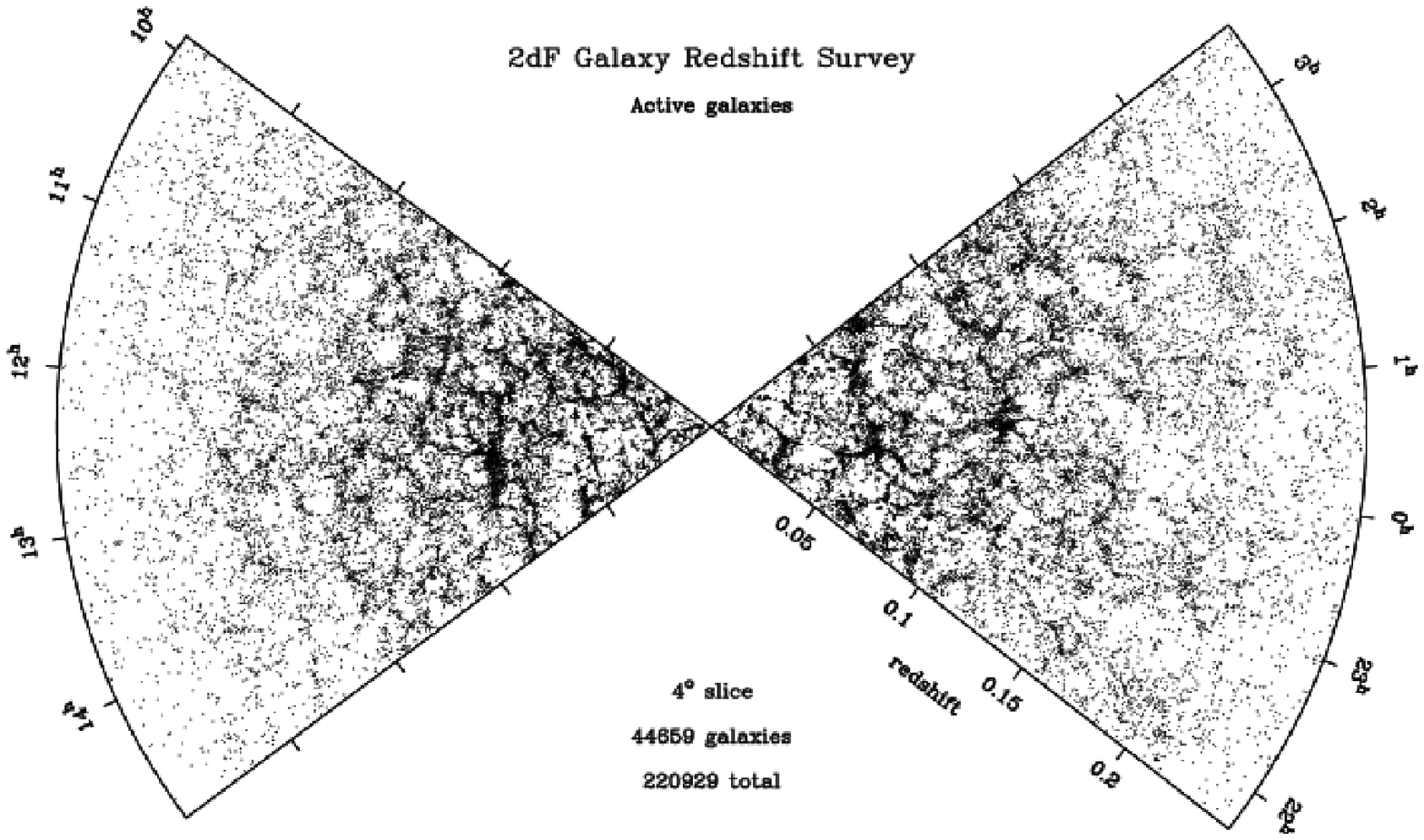} 
}
\hbox{
\epsfxsize=0.9\textwidth
\epsfbox{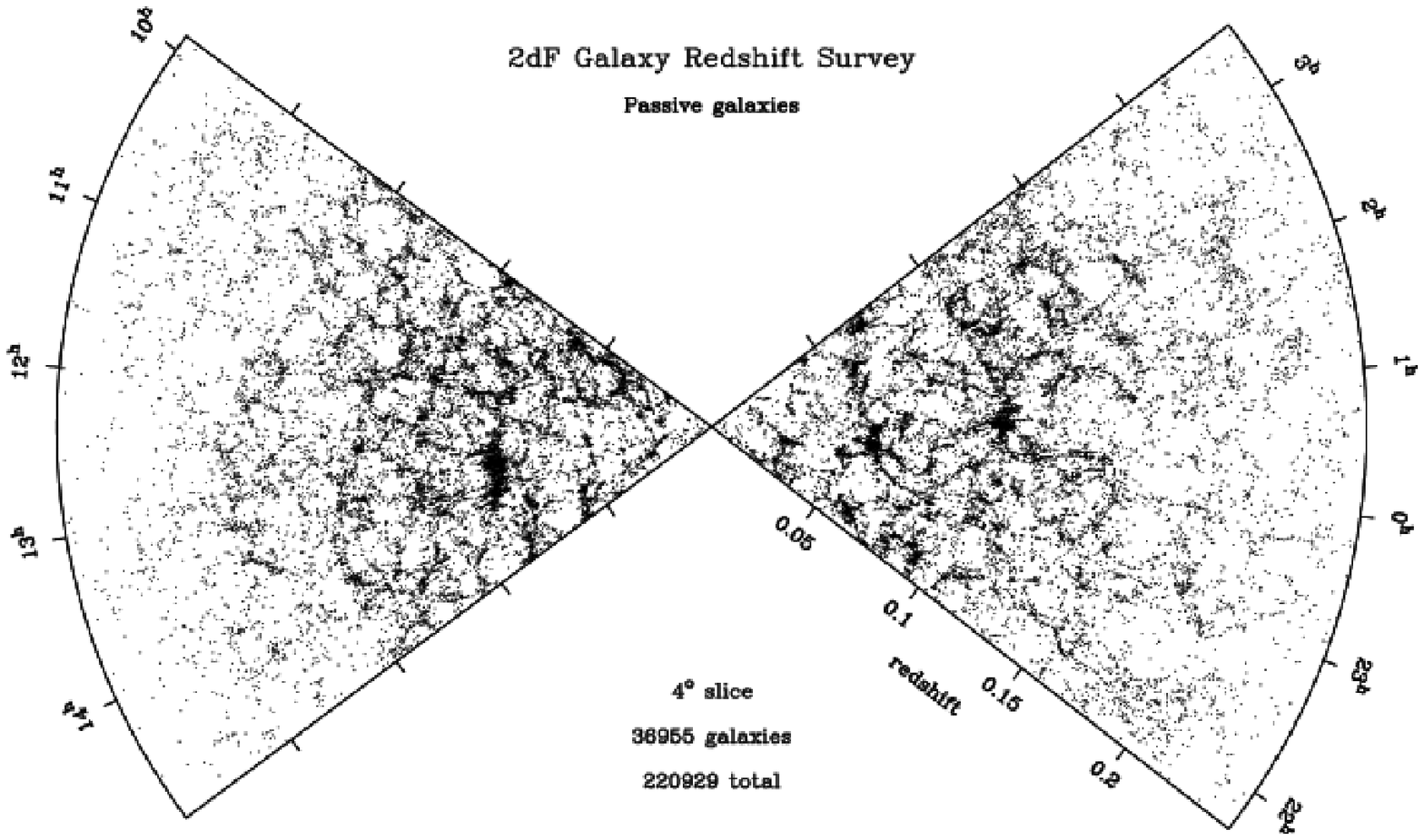}
}
}
\caption{The distribution of galaxies in part of the 2dFGRS, drawn from 
a total of 221,496 galaxies:
slices $4^\circ$ thick, centred at declination
$-2.5^\circ$ in the NGP and $-27.5^\circ$ in the SGP.
The survey is divided at a rest-frame colour of
photographic $B-R=0.85$, into galaxies with and
without active star formation. The 
This image reveals a wealth of detail, including
linear supercluster features, often nearly perpendicular
to the line of sight. It appears that these transverse features
have been enhanced by infall velocities.}
\end{figure}

\sec{Redshift-space correlations}

The simplest statistic for studying clustering in the galaxy
distribution is the the two-point correlation function,
$\xi(\sigma,\pi)$. This measures the excess probability over random of
finding a pair of galaxies with a separation in the plane of the sky
$\sigma$ and a line-of-sight separation $\pi$. Because the radial
separation in redshift space includes the peculiar velocity as well as
the spatial separation, $\xi(\sigma,\pi)$ will be anisotropic. On small
scales the correlation function is extended in the radial direction due
to the large peculiar velocities in non-linear structures such as groups
and clusters -- this is the well-known `Finger-of-God' effect. On large
scales it is compressed in the radial direction due to the coherent
infall of galaxies onto mass concentrations -- the Kaiser effect (Kaiser
1987).

\begin{figure}[ht]
\plotter{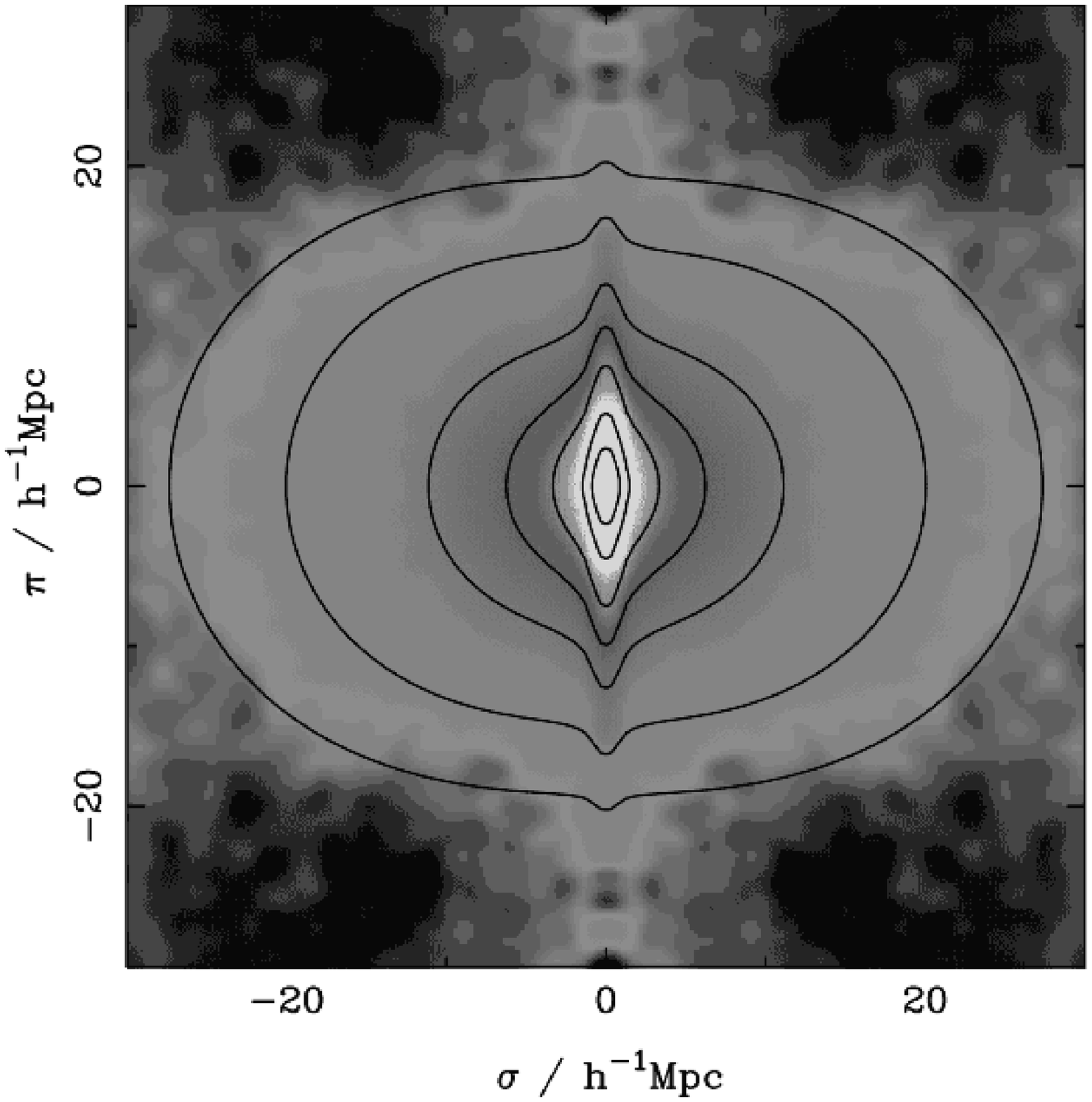}{0.45}
\caption{The galaxy correlation function $\xi(\sigma,\pi)$ as a function
of transverse ($\sigma$) and radial ($\pi$) pair separation is shown as
a greyscale image. It was computed in $0.2\mpcoh$ boxes and then smoothed
with a Gaussian having an rms of $0.5\mpcoh$. The contours are for a model
with $\beta=0.4$ and $\sigma_p=400\kms$, and are plotted at $\xi=10$, 5,
2, 1, 0.5, 0.2 and 0.1.}
\end{figure}

To estimate $\xi(\sigma,\pi)$ we compare the observed count of galaxy
pairs with the count estimated from a random distribution following the
same selection function both on the sky and in redshift
as the observed galaxies. We apply optimal weighting to
minimise the uncertainties due to cosmic variance and Poisson noise.
The redshift-space correlation function for the 2dFGRS 
computed in this way is shown in Figure~5.
The correlation-function results display very clearly
two signatures of redshift-space distortions.
The `fingers of God' from small-scale random
velocities are very clear, as indeed has been the case
from the first redshift surveys (e.g. Davis \& Peebles 1983).
However, this is the first time that the 
large-scale flattening from coherent infall has been seen in detail.
An initial analysis of this effect was performed in Peacock
et al. (2001), and the final database was analysed by
Hawkins et al. (2002).

The degree of large-scale flattening is
determined by the total mass density parameter, $\Omega_m$, and the
biasing of the galaxy distribution.
On large scales, it should be correct to assume a linear bias model,
with correlation functions $\xi_g(r) = b^2 \xi(r)$,
so that the redshift-space distortion on large scales depends on
the combination $\beta \equiv \Omega_m^{0.6}/b$. 
This is modified by the Finger-of-God effect, which is
significant even at large scales and dominant at small scales. 
The effect can be modelled by introducing a parameter
$\sigma_p$, which represents the rms pairwise velocity dispersion of
the galaxies in collapsed structures, $\sigma_p$ (see e.g. Ballinger et al. 1996).
Considering both these effects, and marginalising over $\sigma_p$, the best estimate of
$\beta$ and its 68\% confidence interval according to Hawkins et al. (2002) is
$$
\beta=0.49\pm0.09
$$
The quoted error is slightly larger than in Peacock et al. (2001): mainly,
this reflects the decision of Hawkins et al. to concentrate on the
better sampled volume at $z<0.2$, although a more detailed comparison
with mock data also indicates that the previous errors were too small by
a factor of about 1.2.

Our measurement of $\Omega^{0.6}/b$ can only be used to determine $\Omega$ if
the bias is known. 
We discuss below two methods by which the
bias parameter may be inferred, which in fact favour a
low degree of bias. Nevertheless, there are other uncertainties
in converting a measurement of $\beta$ to a figure for $\Omega$.
The 2dFGRS has a median redshift of 0.11; with weighting,
the mean redshift in Hawkins et al. is 0.15,
and our measurement should be interpreted as $\beta$ at
that epoch. 
The optimal weighting also means that our mean
luminosity is high: it is approximately
1.4 times the characteristic luminosity, $L^*$,  of the overall
galaxy population (Folkes et al. 1999; Madgwick et al. 2002). 
This means that we need to quantify the luminosity dependence
of clustering.

\sec{Real-space clustering and its dependence on luminosity}

The dependence of galaxy clustering on luminosity is an
effect that was controversial for a number of years.
Using the APM-Stromlo redshift survey, Loveday et al. (1995)
claimed that there was no trend of clustering amplitude
with luminosity, except possibly at the very lowest
luminosities. In contradiction, the SSRS study of
Benoist et al. (1996) suggested that the strength of
galaxy clustering increased monotonically with luminosity, with
a particularly marked effect for galaxies above $L^*$.
The latter result was arguably more plausible, based on
what we know of luminosity functions and morphological
segregation. It has been clear for many years that elliptical
galaxies display a higher correlation amplitude than
spirals (Davis \& Geller 1976). Since ellipticals are also
more luminous on average, as shown above,
some trend with luminosity is to
be expected, but the challenge is to detect it.

\begin{figure}[ht]
\plotter{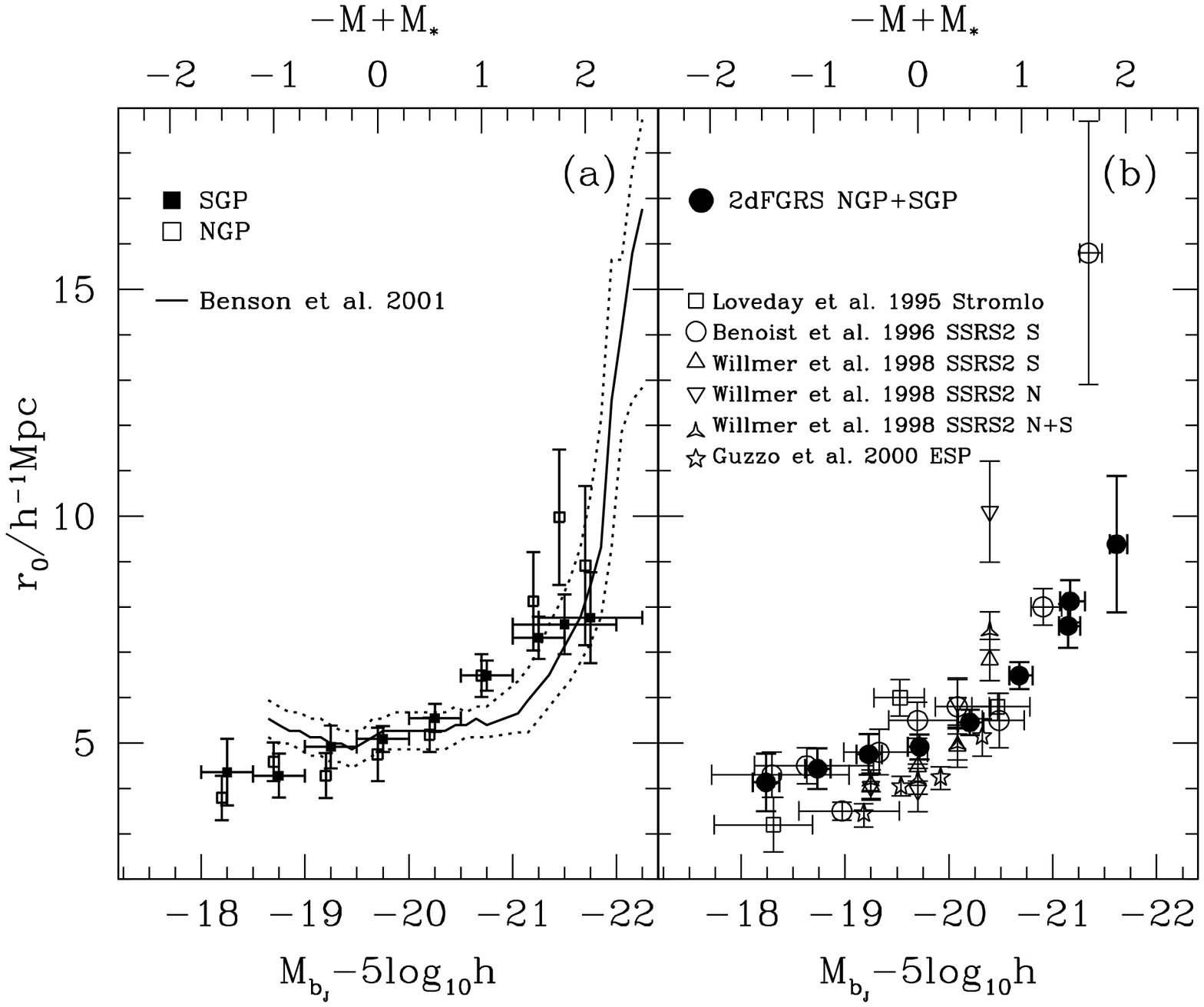}{0.6}
\caption{(a) The correlation length in real space as a function of 
absolute magnitude. 
The solid line shows the predictions of the semi-analytic 
model of Benson et al. (2001), computed in a series of overlapping 
bins, each $0.5$ magnitudes wide. The dotted curves show an 
estimate of the errors on this prediction, including the relevant sample variance
for the survey volume.
(b) The real space correlation length estimated combining 
the NGP and SGP (filled circles). 
The open symbols show a selection of recent data from other studies.
}
\end{figure}

The difficulty with measuring the dependence of $\xi(r)$ on
luminosity is that cosmic variance can mask the signal of
interest. It is therefore important to analyse volume-limited samples
in which galaxies of different luminosities are compared
in the same volume of space. This comparison was undertaken
by Norberg et al. (2001), who measured real-space correlation functions
via the projection $\Xi(\sigma) = \int \xi(\sigma,\pi)\; d\pi$,
demonstrating that it was possible to
obtain consistent results in both NGP and SGP.
A very clear detection of luminosity-dependent clustering was achieved,
as shown in Figure~6. The results
can be described by a linear dependence of effective bias
parameter on luminosity:
$$
b/b^* = 0.85 + 0.15\,(L/L^*),
$$
and the scale-length of the real-space correlation function for $L^*$
galaxies is approximately $r_0=4.8 \mpcoh$. 
This trend is in qualitative agreement with the results of
Benoist et al. (1996), but in fact these workers gave a stronger
dependence on luminosity than is indicated by the 2dFGRS.
Finally, with spectral classifications, it is possible to
measure the dependence of clustering both on luminosity and
on spectral  type, to see to what extent morphological
segregation is responsible for this result. Norberg et al. (2002)
show that, in fact, the principal effect seems to be with
luminosity: $\xi(r)$ increases with $L$ for all spectral types.
This is reasonable from a theoretical point of view, in which
the principal cause of different clustering amplitudes is the
mass of halo that hosts a galaxy 
(e.g. Cole \& Kaiser 1989; Mo \& White 1996; 
Kauffman, Nusser \& Steinmetz 1997).

Finally, these results would lead us to infer that the LF
must change in strongly clumped regions, shifting
to higher luminosities. Such an effect has been sought for many years,
but always yielded null results. However, De Propris et al.
(2002) have shown that $L^*$ in rich clusters does
obey a shift with respect to the global value, being
brighter by $0.28\pm0.08$ mag.

\sec{The 2dFGRS power spectrum}

Perhaps the key aim of the 2dFGRS was to perform an accurate
measurement of the 3D clustering power spectrum, in order
to improve on the APM result,
which was deduced by deprojection of angular
clustering (Baugh \& Efstathiou 1993, 1994). 
The results of this direct estimation of the 3D power
spectrum are shown in Figure~7.
This power-spectrum estimate uses the FFT-based approach
of Feldman, Kaiser \& Peacock (1994), and needs to be interpreted
with care. Firstly, it is a raw redshift-space estimate, so
that the power beyond $k\simeq 0.2 \hompc$ is severely damped
by fingers of God. On large scales, the power is enhanced, both
by the Kaiser effect and by the luminosity-dependent clustering
discussed above. Finally, the FKP estimator yields the
true power convolved with the window function. This
modifies the power significantly on large scales (roughly
a 20\% correction). We have made an approximate correction for
this in Figure~7. The precision of the
power measurement appears to be encouragingly high, and the
next task is to perform a detailed fit of physical
power spectra, taking full account of the window effects.
We summarize here results from this analysis
(Percival et al. 2001).

\begin{figure}[ht]
\plotter{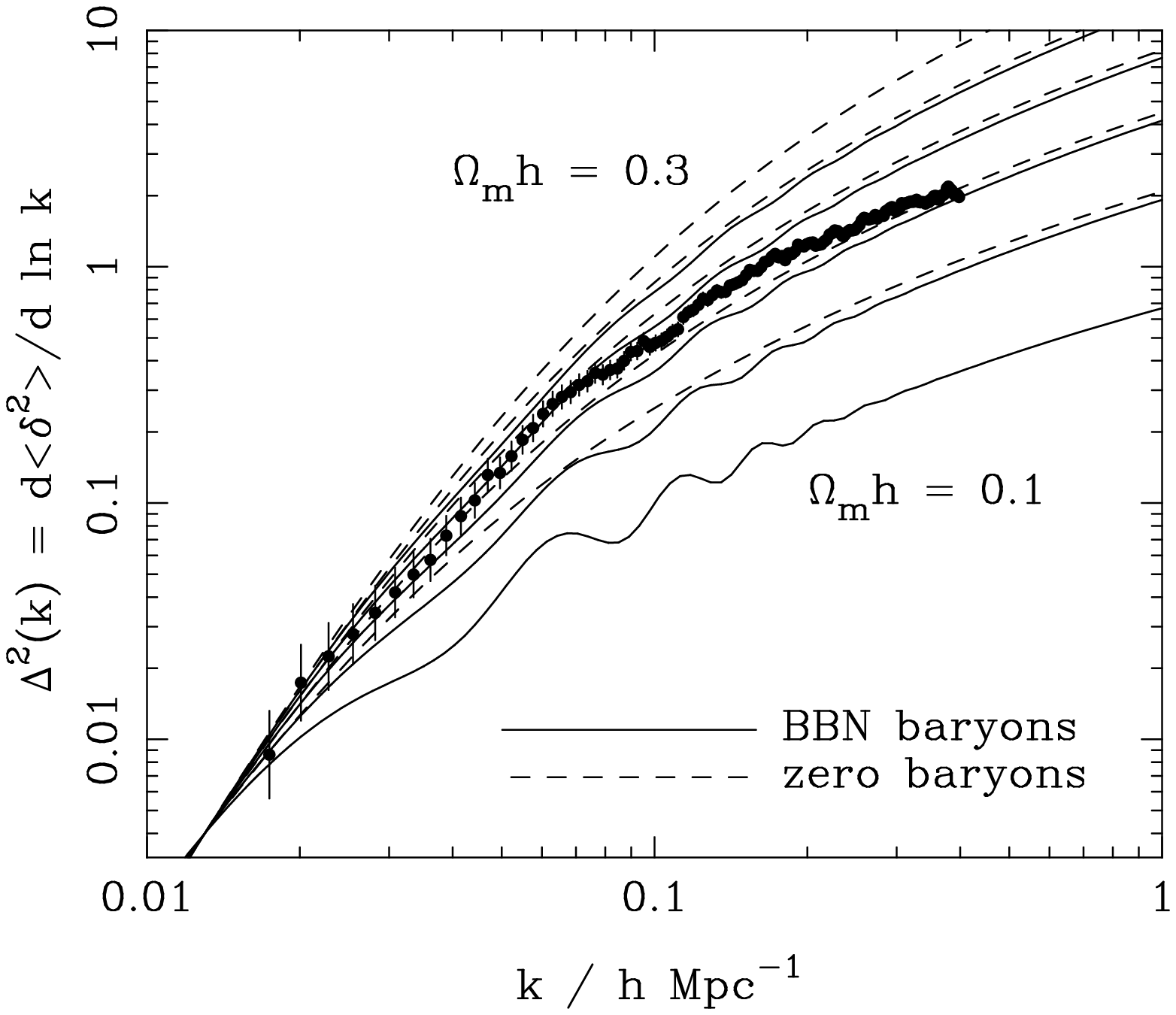}{0.55}
\caption{The 2dFGRS redshift-space dimensionless power spectrum, 
$\Delta^2(k)$,
estimated according to the FKP procedure. The solid points
with error bars show the power estimate. The window
function correlates the results at different $k$ values,
and also distorts the large-scale shape of the power spectrum
An approximate correction for the latter effect has been applied.
The solid and dashed lines show various CDM models, all assuming
$n=1$. For the case with non-negligible baryon content,
a big-bang nucleosynthesis value of $\Omega_b h^2=0.02$ is
assumed, together with $h=0.7$. A good fit is clearly obtained
for $\Omega_m h \simeq 0.2$. Note that the observed power at
large $k$ will be boosted by nonlinear effects, but damped by 
small-scale random peculiar velocities. It appears that these
two effects very nearly cancel, but model fitting is generally
performed only at $k<0.15 \hompc$ in order to avoid these complications.}
\end{figure}

The fundamental assumption is that, on large scales, linear biasing
applies, so that the nonlinear galaxy power spectrum in redshift space has a shape
identical to that ow linear theory in real space.
We believe that this assumption is valid for $k<0.15\hompc$;
the detailed justification comes from analyzing realistic 
mock data derived from $N$-body simulations (Cole et al 1998).
The model free parameters are thus the primordial spectral
index, $n$, the Hubble parameter, $h$, the total matter
density, $\Omega_m$, and the baryon fraction, $\Omega_b/\Omega_m$.
Note that the vacuum energy does not enter. Initially, we
show results assuming $n=1$; this assumption is relaxed later.

In order to compare the 2dFGRS power spectrum to members of the
CDM family of theoretical models, it is essential to have a
proper understanding of the full covariance matrix of the data:
the convolving effect of the window function 
causes the power at adjacent $k$ values to be correlated.
This covariance matrix was estimated by applying the survey window to a
library of Gaussian realisations of linear density fields.
Similar results were obtained using a covariance matrix
estimated from a set of mock catalogues.
It is now possible to explore the space of CDM models, and
likelihood contours in $\Omega_b/\Omega_m$ versus $\Omega_mh$
are shown in Figure~8. At each point in this
surface we have marginalized by integrating the likelihood surface
over the two free parameters, $h$ and the power spectrum
amplitude. 
Assuming a uniform prior for $h$ over a factor of 2 is arguably
over-cautious, and we have therefore added a Gaussian prior $h=0.7\pm
10\%$. This corresponds to multiplying by
the likelihood from external constraints such as the HST key project
(Freedman et al. 2001); this has only a minor effect on the results.

\begin{figure}[ht]
\japplottwo{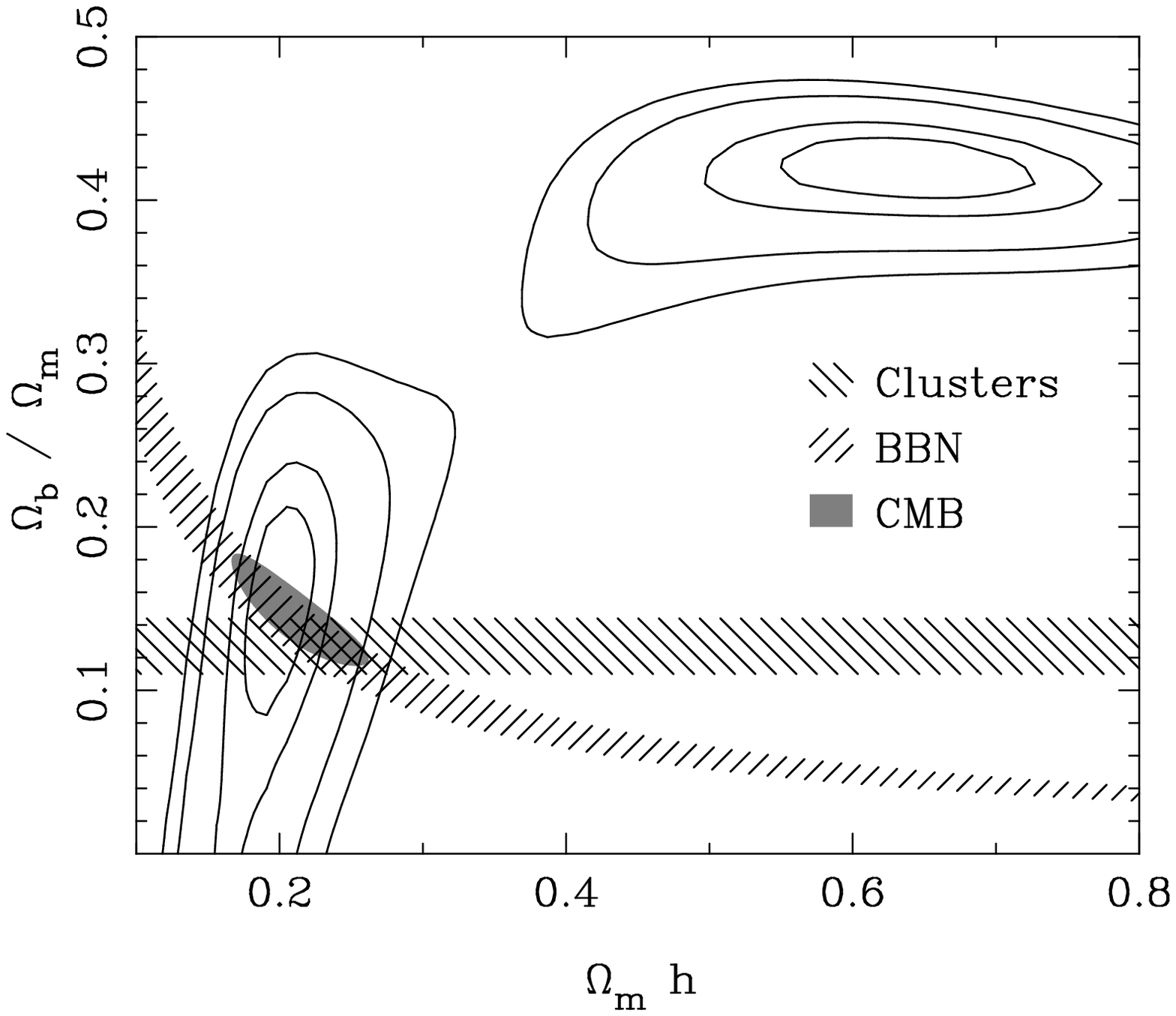}{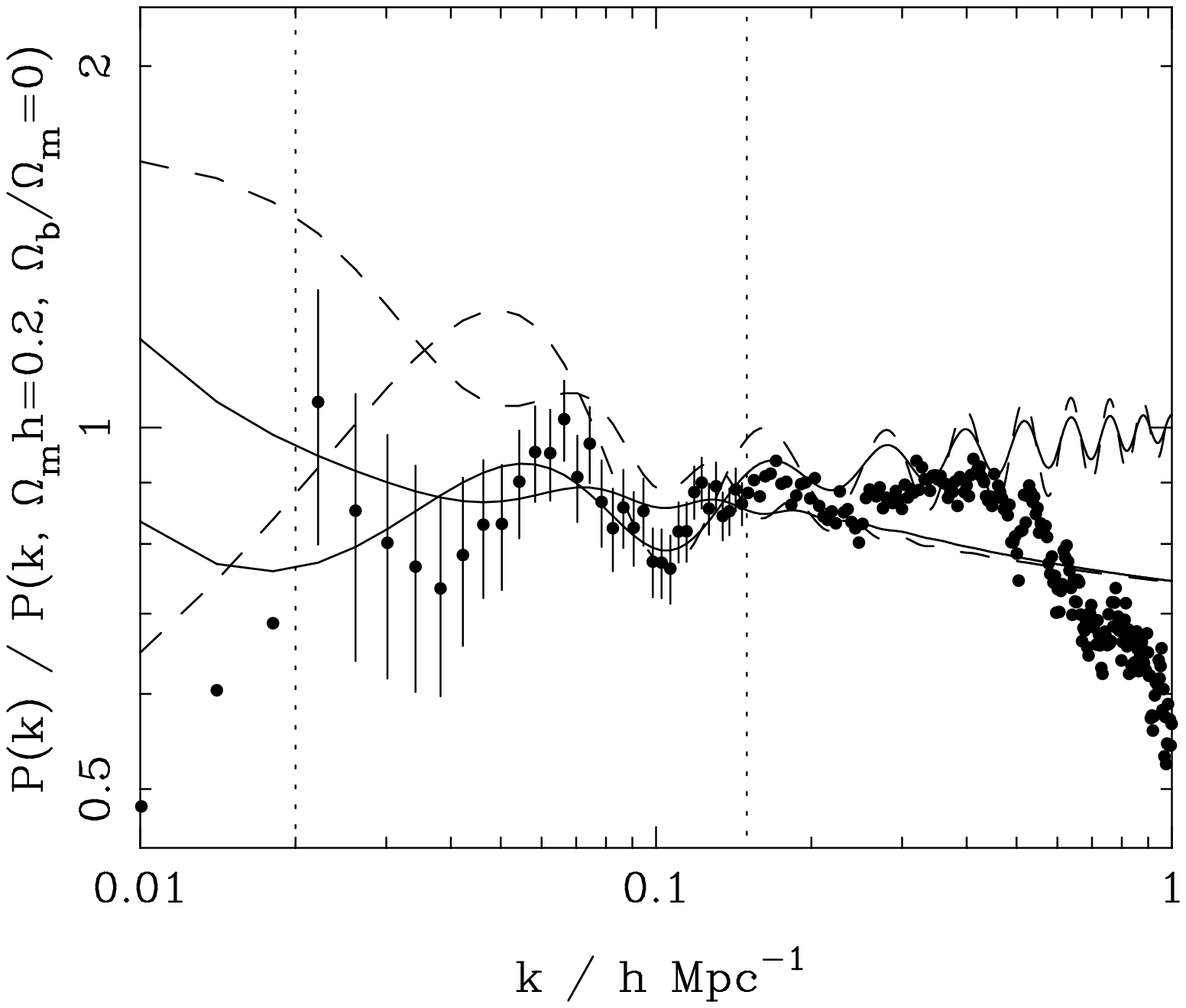}
\caption{Likelihood contours for the best-fit linear power spectrum
  over the region $0.02<k<0.15$. The normalization is a free parameter
  to account for the unknown large scale biasing. Contours are plotted
  at the usual positions for one-parameter confidence of 68\%, and
  two-parameter confidence of 68\%, 95\% and 99\% (i.e. $-2\ln({\cal
  L}/{\cal L_{\rm max}}) = 1, 2.3, 6.0, 9.2$). We have marginalized
  over the missing free parameters ($h$ and the power spectrum
  amplitude).
  A prior on $h$ of $h=0.7\pm 10\%$ was assumed. 
  This result is compared to estimates from X-ray cluster
  analysis (Evrard 1997), big-bang nucleosynthesis (Burles et al.
  2001) and CMB results (Netterfield et al. 2001; Pryke et al. 2002). 
  The CMB results assume that
  $\Omega_bh^2$ and $\Omega_{\rm cdm}h^2$ were independently
  determined from the data.
The second panel shows the 2dFGRS data compared with the two preferred models from
  the Maximum Likelihood fits convolved with the window function
  (solid lines). Error bars show the diagonal elements of the
  covariance matrix, for the fitted data that lie between the dotted
  vertical lines. The unconvolved models are also shown (dashed
  lines). The $\Omega_m h \simeq 0.6$, $\Omega_b/\Omega_m=0.42$,
  $h=0.7$ model has the higher bump at $k\simeq 0.05\hompc$. The
  smoother $\Omega_m h \simeq 0.20$, $\Omega_b/\Omega_m=0.15$, $h=0.7$
  model is a better fit to the data because of the overall shape.
A preliminary analysis of the complete final 2dFGRS sample yields
a slightly smoother spectrum than the results shown
here (from Percival et al. 2001), so that the high-baryon solution becomes
disfavoured.
}
\end{figure}

Figure~8 shows that there is a degeneracy between
$\Omega_mh$ and the baryonic fraction $\Omega_b/\Omega_m$. However, there
are two local maxima in the likelihood, one with $\Omega_mh \simeq 0.2$
and $\sim 20\%$ baryons, plus a secondary solution $\Omega_mh \simeq 0.6$
and $\sim 40\%$ baryons. The high-density model can be rejected through a variety
of arguments, and the preferred solution is
$$
  \Omega_m h = 0.20 \pm 0.03; \quad\quad \Omega_b/\Omega_m = 0.15 \pm 0.07.
$$
The 2dFGRS data are compared to the best-fit linear power spectra
convolved with the window function in Figure~8. This
shows where the two branches of solutions come from: the low-density
model fits the overall shape of the spectrum with relatively small
`wiggles', while the solution at $\Omega_m h \simeq 0.6$ provides a
better fit to the bump at $k\simeq 0.065\hompc$, but fits the overall
shape less well.
A preliminary analysis of $P(k)$ from the full final dataset
shows that $P(k)$ becomes smoother: 
the high-baryon solution becomes disfavoured, and
the uncertainties narrow slightly around the lower-density solution:
$\Omega_m h = 0.18 \pm 0.02$; $\Omega_b/\Omega_m = 0.17 \pm 0.06$.

It is interesting to compare these conclusions with other
constraints. These are shown on Figure~8, assuming 
$h=0.7\pm 10\%$.
Latest estimates of the Deuterium to Hydrogen ratio in QSO spectra
combined with big-bang nucleosynthesis theory predict $\Omega_bh^2 =
0.020\pm 0.001$ (Burles et al. 2001), which translates to the
shown locus of $f_{\japsub B}$ vs $\Omega_m h$. X-ray
cluster analysis predicts a baryon fraction
$\Omega_b/\Omega_m=0.127\pm0.017$ (Evrard 1997) which is within
$1\sigma$ of our value. These loci intersect very close
to our preferred model. Moreover, 
these results are in good agreement with independent
estimates of the total density and baryon content from
data on CMB anisotropies
(e.g. Netterfield et al. 2001; Pryke et al. 2002).

Perhaps the main point to emphasise here is that the 2dFGRS results are not
greatly sensitive to the assumed tilt of the primordial spectrum. We
have used CMB results to motivate the choice of $n=1$, 
as discussed below, but it is
clear that very substantial tilts are required to alter the
conclusions significantly: $n\simeq 0.8$ would be required to turn
zero baryons into the preferred model.

The main residual worry about accepting the above conclusions is 
probably whether the assumption of linear bias can really be valid. 
In general, concentration towards higher-density regions both
raises the amplitude of clustering, but also steepens the correlations,
so that bias is largest on small scales. One way in which
this issue can be studied is to use the split by colour introduced
above. Figure~9 shows the power spectra for the 2dFGRS divided in this
way. The shapes are almost identical (perhaps not so surprising,
since the cosmic variance effects are closely correlated in these
co-spatial samples). However, what is impressive is that the
relative bias is almost precisely independent of scale,
even though the passive subset is rather strongly biased
relative to the active subset (relative $b\simeq 1.4$). This provides some
reassurance that the large-scale $P(k)$ reflects the underlying
properties of the dark matter, 
rather than depending on the particular class
of galaxies used to measure it.

\begin{figure}[ht]
\japplottwo{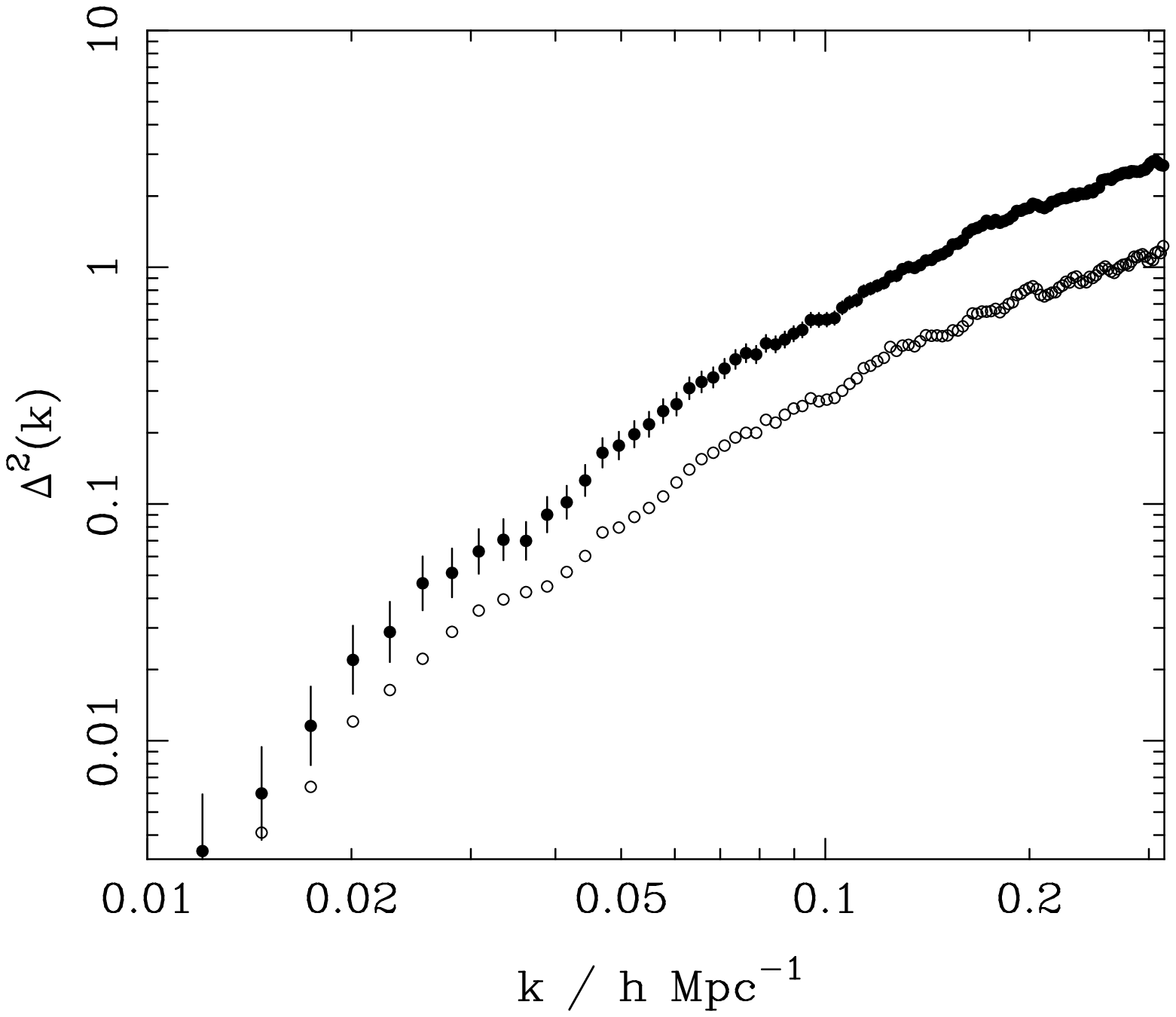}{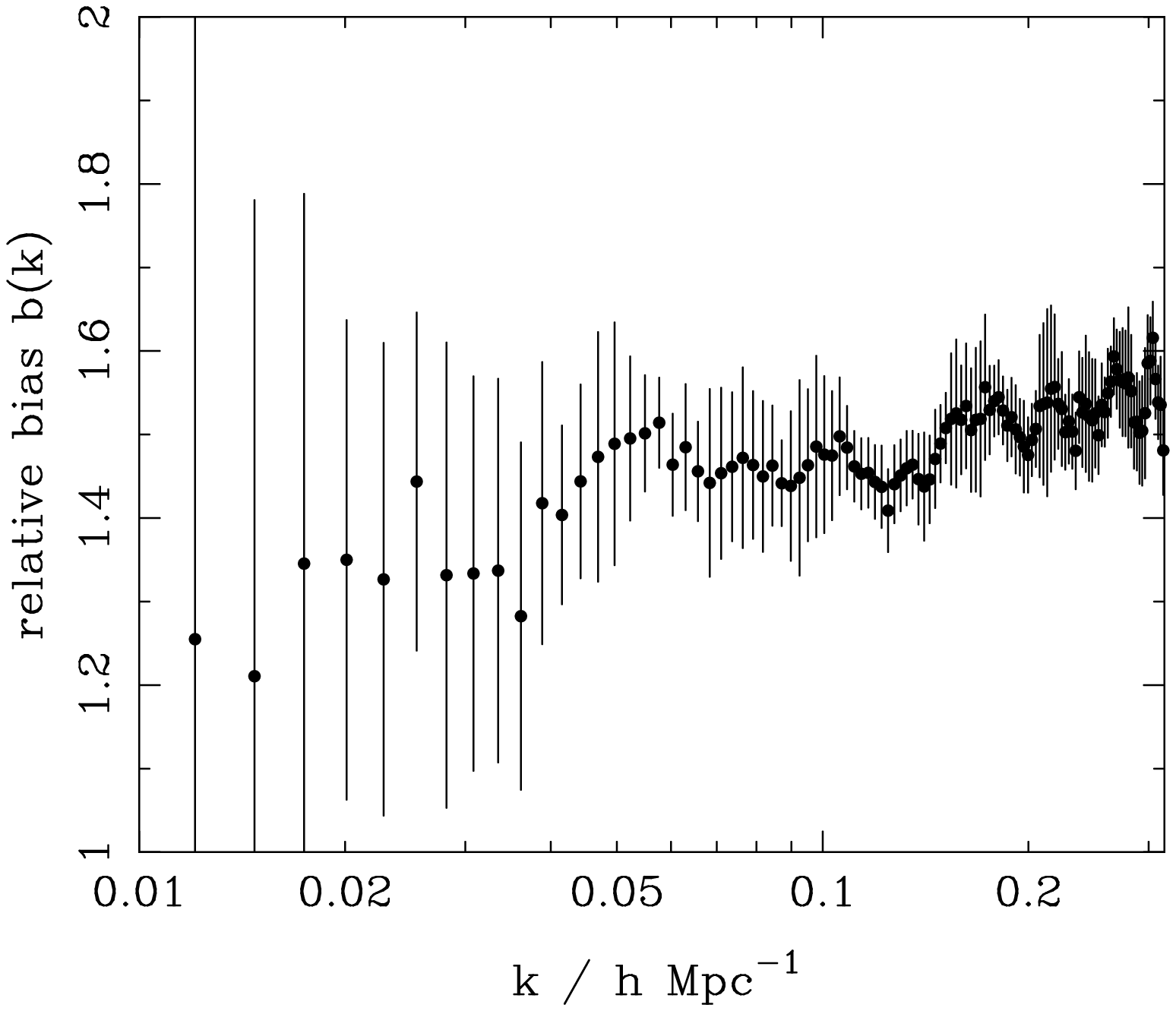}
\caption{
The power spectra of passive galaxies (filled circles) and active galaxies (open circles).
The shapes are strikingly similar.
The square root of the ratio yields the right-hand panel:
the relative bias in redshift space of passive and active galaxies.
The error bars are obtained by a jack-knife analysis. The relative
bias is consistent with a constant value of 1.4 over the range used for
fitting of the power-spectrum data ($0.015 < k < 0.15 \hompc$).
}
\end{figure}

\sec{Combination with the CMB and cosmological parameters}

The 2dFGRS power spectrum contains important information about the
key parameters of the cosmological model, but we have seen that additional
assumptions are needed, in particular the values of $n$ and $h$.
Observations of CMB
anisotropies can in principle measure 
most of the cosmological parameters, and
combination with the 2dFGRS can lift most of the degeneracies inherent
in the CMB-only analysis. It is therefore of interest to see
what emerges from a joint analysis.

These issues are discussed in Efstathiou et al. (2002).
The CMB data alone contain two important degeneracies:
the `geometrical' and `tensor' degeneracies.
In the former case, one can evade the
commonly-stated CMB conclusion that the
universe is flat, by adjusting both $\Lambda$ and $h$ to
extreme values. In the latter case, a model with a large
tensor component can be made to resemble a zero-tensor model
with large blue tilt ($n>1$) and high baryon content.
Efstathiou et al. (2002) show that adding the 2dFGRS
data removes the first degeneracy, but not the second.
This is reasonable: if we take the view that the CMB determines
the physical density $\Omega_m h^2$, then a measurement of
$\Omega_m h$ from 2dFGRS gives both
$\Omega_m$ and $h$ separately in principle,
removing one of the degrees of freedom on which the geometrical
degeneracy depends. On the other hand, the 2dFGRS alone constrains
the baryon content weakly, so this does not remove the scope
for the tensor degeneracy.

On the basis of this analysis, we can therefore be confident
that the universe is very nearly flat
($|\Omega-1|<0.05$), so it is defensible to assume
hereafter that this is exactly true. The importance of tensors
will of course be one of the key questions for cosmology over the
next several years, but it is interesting to consider the limit
in which these are negligible. In this case, the standard model
for structure formation contains a vector of only 6 parameters:
$$
{\bf p} = (n_s, \Omega_m, \Omega_b, h, Q, \tau).
$$
Of these, the optical depth to last scattering, $\tau$, is almost entirely
degenerate with the normalization, $Q$ -- and indeed with
the bias parameter; we discuss this below.
The remaining four parameters are pinned down very precisely:
using the latest CMB data plus the 2dFRGS power spectrum,
we obtain
$$
(n_s, \Omega_c h^2, \Omega_b h^2, h) =
(0.963\pm 0.042, 0.115\pm 0.009, 0.021\pm0.002, 0.665 \pm 0.047),
$$
or an overall density parameter of $\Omega_m=0.31 \pm 0.05$.

It is remarkable how well these figures agree with completely
independent determinations: $h=0.72\pm 0.08$ from the HST key project
(Mould et al. 2000; Freedman et al. 2001);
$\Omega_b h^2 =0.020 \pm 0.001$ (Burles et al. 2001).
This gives confidence that the tensor component must
indeed be sub-dominant.
For further details of this analysis, see Percival et al. (2002).

\sec{Matter fluctuation amplitude and bias}

The above conclusions were obtained by considering the shapes
of the CMB and galaxy power spectra. However, it is also of
great interest to consider the amplitude of mass fluctuations,
since a comparison with the galaxy power spectrum
allows us to infer the degree of bias directly.
This analysis was performed by Lahav et al. (2002).
Given assumed values for the cosmological parameters, the
present-day linear normalization of the mass spectrum (e.g. $\sigma_8$)
can be inferred.
It is convenient to define a corresponding measure
for the galaxies, $\sigma_{8{\rm g}}$, such that we can express the bias parameter
as
$$
b = {\sigma_{8{\rm g}} \over \sigma_{8{\rm m}} }.
$$
In practice, we define $\sigma_{8{\rm g}}$ to be the value  required
to fit a CDM model to the power-spectrum data on linear scales ($0.02<k<0.15 \hompc$).
A final necessary complication of the notation is that we need to distinguish
between the apparent values 
of $\sigma_{8{\rm g}}$ as measured in redshift space
($\smash{\sigma_{8{\rm g}}^S}$) and the real-space value that would be measured in the
absence of redshift-space distortions ($\smash{\sigma_{8{\rm g}}^R}$). It is the latter
value that is required in order to estimate the bias.

A model grid covering 
the range
$0.1 < \Omega_m h < 0.3 $, $0.0 < \Omega_b/\Omega_m < 0.4 $, 
$0.4 < h < 0.9$  and $0.75 < \smash{\sigma^S_{8{\rm g}}} < 1.14$
was considered. The primordial index was assumed to be
$n=1$ initially, and the dependence on $n$ studied separately.
For fixed 
`concordance model' parameters
$n=1$, $k=0$, $\Omega_m = 0.3$, $\Omega_b h^2 = 0.02$
and a Hubble constant $h=0.70$, 
we find that the amplitude of 2dFGRS galaxies
in redshift space is $\smash{\sigma_{8{\rm g}}^S} (L_s,z_s) = 0.94$.
Correcting for redshift-space distortions as detailed above
reduces this to 0.86 in real space. Applying a correction for
a mean luminosity of $1.9L^*$ using the recipe of Norberg et al. (2001),
we obtain an estimate of $\smash{\sigma_{8{\rm g}}^R} (L^*,z_s) = 0.76$,
with a negligibly small random error.
In order to obtain present-day bias figures, we need to know
the evolution of galaxy clustering to $z=0$. Existing data on
clustering evolution reveals very slow changes: higher bias at early
times largely cancels the evolution of the dark matter.
We therefore assume no evolution in $\smash{\sigma_{8{\rm g}}}$.

The value of $\sigma_8$ for the dark matter can be deduced from the
CMB fits:
$$
 \sigma_8 = (0.72 \pm 0.04) \, \exp \tau,
$$
where the error bar includes both data errors and theory
uncertainty. The unsatisfactory feature is the degeneracy
with the optical depth to last scattering. For reionization
at redshift $8$, we would have $\tau\simeq 0.05$;
it is unlikely that $\tau$ can be hugely larger
(e.g. Loeb \& Barkana 2001). Although direct removal of this
theoretical prejudice is desirable (and will be possible with
future CMB data), it seems reasonable to assume that
the true value of $\sigma_8$ must be very close to 0.76. Within the errors, this
agrees perfectly with our $\smash{\sigma_{8{\rm g}}^R} (L^*,0) = 0.76$,
implying that $L^*$ galaxies are very nearly exactly unbiased.
As we have seen, there are large variations
in the clustering amplitude with type, so that this outcome
must be something of a coincidence.

Finally, this conclusion of near-unity bias was reinforced
in a completely independent way, by using the
measurements of the bispectrum of galaxies in the 2dFGRS
(Verde et al. 2002). As it is based on three-point
correlations, this statistic is sensitive to the filamentary
nature of the galaxy distribution -- which is a signature of
nonlinear evolution. One can therefore split the degeneracy between
the amplitude of dark-matter fluctuations and the
amount of bias. At the effective redshift and luminosity
of their sample ($z_s=0.17$ and $L=1.9 L^*$), Verde et al.
found $b=1.04\pm 0.11$. Although the corrections to zero
redshift and to luminosity $L^*$ are probably significant,
this reinforces the point that on large scales there is
no substantial difference in clustering between typical
galaxies and the dark matter (small scales, of course,
are another matter entirely).

\sec{Conclusions}

The 2dFGRS is the first 3D survey of the local universe
to achieve 100,000 redshifts, almost an order of
magnitude improvement on previous work.
The final database should yield definitive results on a
number of key issues relating to galaxy clustering.
For details of the current status of the 2dFGRS, see 
{\tt http://www.mso.anu.edu.au/2dFGRS}.
In particular, this site gives details of the 2dFGRS
public release policy, in which 
approximately the first half of the survey data were made available
in June 2001, with the complete survey database to be made
public by mid-2003. Some key results of the survey to date
may be summarized as follows:

\japitem{(1)}The galaxy luminosity function has been measured
precisely as a function of spectral type (Folkes et al. 1999;
Madgwick et al. 2002).

\japitem{(2)}The amplitude of galaxy clustering has been shown
to depend on luminosity (Norberg et al. 2001). The relative
bias is $b/b^* = 0.85 + 0.15\, (L/L^*)$.

\japitem{(3)}The redshift-space correlation function has been
measured out to $30\mpcoh$. Redshift-space distortions imply
$\beta \equiv \Omega_m^{0.6}/b = 0.49 \pm 0.09$, for
galaxies with $L\simeq 1.4 L^*$.

\japitem{(4)}The galaxy power
spectrum has been measured to high accuracy (10--15\% rms) over about a
decade in scale at $k<0.15\hompc$. The results are very well matched by
an $n=1$ CDM model with $\Omega_mh=0.18$ and 16\% baryons.

\japitem{(5)}Combining the power spectrum results with current CMB data,
very tight constraints are obtained on cosmological parameters. 
For a scalar-dominated flat model, we obtain $\Omega_m=0.31\pm 0.05$,
and $h=0.68 \pm 0.04$, independent of external data.

\japitem{(6)}Results from the CMB comparison 
imply a large-scale bias parameter consistent with unity.
This conclusion is also reached in a completely independent way
via the bispectrum analysis of Verde et al. (2002).

\enditem
Overall, these results provide precise support for a cosmological
model that is flat, with $(\Omega_b,\Omega_c,\Omega_v)
\simeq (0.04,0.25,0.71)$, to a tolerance of 10\% in each figure.
Although the $\Lambda$CDM model has been claimed to have problems in matching
galaxy-scale observations, it clearly works extremely well on large scales, and
any proposed replacement for CDM will have to maintain this agreement.
So far, there has been no need to invoke either tilt of
the scalar spectrum, or a tensor component in the CMB.
If this situation is to change, the most likely route will be
via new CMB data,  combined with the key complementary information
that the large-scale structure in the 2dFGRS can provide.

\section*{Acknowledgements}

The 2dF Galaxy Redshift Survey
was made possible by the dedicated efforts of the staff
of the Anglo-Australian Observatory, both in creating the 2dF
instrument, and in supporting it on the telescope.


\section*{References}

\japref Ballinger W.E., Peacock J.A., Heavens A.F., 1996, MNRAS, 282, 877
\japref Baugh C.M., Efstathiou G., 1993, MNRAS, 265, 145
\japref Baugh C.M., Efstathiou G., 1994, MNRAS, 267, 323
\japref Benoist C., Maurogordato S., da Costa L.N., Cappi A., Schaeffer R., 1996, ApJ, 472, 452
\japref Benson, A.J., Frenk, C.S., Baugh, C.M., Cole, S., Lacey, C.G., 2001, MNRAS, 327, 1041
\japref Burles S., Nollett K.M., Turner M.S., 2001, ApJ, 552, L1
\japref Cole S., Kaiser N., 1989, MNRAS, 237, 1127
\japref Cole S., Hatton S., Weinberg D.H., Frenk C.S., 1998, MNRAS, 300, 945
\japref Colless M. et al., 2001, MNRAS, 328, 1039
\japref Davis M., Geller M.J., 1976, ApJ, 208, 13
\japref Davis M., Peebles, P.J.E., 1983, ApJ, 267, 465
\japref De Propris R. et al., 2002, astro-ph/0212562
\japref Efstathiou G. et al., 2002, MNRAS, 330, L29
\japref Evrard A., 1997, MNRAS, 292, 289
\japref Folkes S.J. et al., 1999, MNRAS, 308, 459
\japref Feldman H.A., Kaiser N., Peacock J.A., 1994, ApJ, 426, 23
\japref Freedman W.L. et al., 2001, ApJ, 553, 47 
\japref Hambly N.C., Irwin M.J., MacGillivray H.T., 2001, MNRAS 326 1295
\japref Hawkins E.J. et al., 2002, astro-ph/0212375
\japref Kaiser N., 1987, MNRAS, 227, 1
\japref Kauffmann G., Nusser A., Steinmetz M., 1997, MNRAS, 286, 795
\japref Lahav O. et al., 2002, MNRAS, 333, 961
\japref Lewis I.J. et al., 2002, MNRAS, 333, 279
\japref Loeb A., Barkana R., 2001, ARAA, 39, 19
\japref Loveday J., Maddox S.J., Efstathiou G., Peterson B.A., 1995, ApJ, 442, 457
\japref Maddox S.J., Efstathiou G., Sutherland W.J., Loveday J., 1990a, MNRAS, 242, 43{\sc p}
\japref Maddox S.J., Sutherland W.J., Efstathiou G., Loveday J., 1990b, MNRAS, 243, 692
\japref Maddox S.J., Efstathiou G., Sutherland W.J., 1990c, MNRAS, 246, 433
\japref Madgwick D. et al., 2002, 333, 133
\japref Mo H.J., White S.D.M., 1996, MNRAS, 282, 347
\japref Mould J.R. et al., 2000, ApJ, 529, 786
\japref Netterfield C.B. et al., 2001, astro-ph/0104460
\japref Norberg P. et al., 2001, MNRAS, 328, 64
\japref Norberg P. et al., 2002, MNRAS, 332, 827
\japref Peacock J.A. et al., 2001, Nature, 410, 169
\japref Percival W.J. et al., 2001, MNRAS, 327, 1297
\japref Percival W.J. et al., 2002, MNRAS, 337, 1068
\japref Pryke C. et al., 2002, ApJ, 568, 46
\japref Schlegel D.J., Finkbeiner D.P., Davis M., 1998, ApJ, 500, 525
\japref Stoughton C.L. et al., 2002, AJ, 123, 485
\japref Verde L. et al., 2002, MNRAS, 335, 432

\end{document}